\documentclass[a4paper,twoside]{article}

\usepackage{epsfig}
\usepackage{subcaption}
\usepackage{calc}
\usepackage{amssymb}
\usepackage{amstext}
\usepackage{amsmath}
\usepackage{amsthm}
\usepackage{multicol}
\usepackage{pslatex}
\usepackage{apalike}
\usepackage{algorithm2e}
\usepackage[bottom]{footmisc}

\usepackage{array}
\usepackage{algorithmic}
\usepackage{amssymb}
\usepackage{textcomp}
\usepackage{xcolor}
\usepackage{listings}
\usepackage{verbatim}
\usepackage{balance}
\usepackage[most]{tcolorbox}
\usepackage{hyperref}
\usepackage{csquotes}
\usepackage{enumitem}
\usepackage{orcidlink}

\usepackage{SCITEPRESS}     

\newcolumntype{R}[1]{>{\raggedleft\let\newline\\\arraybackslash}p{#1}} 
\newcolumntype{L}[1]{>{\raggedright\let\newline\\\arraybackslash}p{#1}} 
\newcolumntype{C}[1]{>{\centering\let\newline\\\arraybackslash}p{#1}}

\MakeOuterQuote{"}

\definecolor{main}{HTML}{808080}    
\definecolor{sub}{HTML}{f2f2f2} 
\tcbset{
	sharp corners,
	colback = white,
	before skip = 0.2cm,    
	after skip = 0.5cm      
}
\newtcolorbox{findingsBox}{
	colback = sub, 
	colframe = main, 
	boxrule = 0pt, 
	leftrule = 6pt 
}

\def\BibTeX{{\rm B\kern-.05em{\sc i\kern-.025em b}\kern-.08em
    T\kern-.1667em\lower.7ex\hbox{E}\kern-.125emX}}

\begin{document}

\title{An Online Integrated Development Environment for Automated Programming Assessment Systems}

\author{\authorname{Eduard Frankford\sup{1}\orcidlink{0009-0005-5959-4936
}, Daniel Crazzolara\sup{1}\orcidlink{0009-0007-3548-4665}, Michael Vierhauser\sup{1}\orcidlink{0000-0003-2672-9230}, Niklas Meißner\sup{2}\orcidlink{0000-0001-9929-1220} \\ Stephan Krusche\sup{3}\orcidlink{0000-0002-4552-644X} and Ruth Breu\sup{1}\orcidlink{0000-0001-7093-4341}}
\affiliation{\sup{1}University of Innsbruck, Department of Computer Science, Austria}
\email{\{eduard.frankford, daniel.crazzolara, michael.vierhauser, ruth.breu\}@uibk.ac.at}
\affiliation{\sup{2}University of Stuttgart, Institute of Software Engineering, Germany}
\email{niklas.meissner@iste.uni-stuttgart.de}
\affiliation{\sup{3}Technical University of Munich, School of Computation, Information and Technology, Germany}
\email{krusche@tum.de}
}


\abstract{The increasing demand for programmers has led to a surge in participants in programming courses, making it increasingly challenging for instructors to assess student code manually. As a result, automated programming assessment systems (APASs) have been developed to streamline this process. These APASs support lecturers by managing and evaluating student programming exercises at scale. However, these tools often do not provide feature-rich online editors compared to their traditional integrated development environments (IDEs) counterparts. This absence of key features, such as syntax highlighting and autocompletion, can negatively impact the learning experience, as these tools are crucial for effective coding practice. To address this gap, this research contributes to the field of programming education by extracting and defining requirements for an online IDE in an educational context and presenting a prototypical implementation of an open-source solution for a scalable and secure online IDE. The usability of the new online IDE was assessed using the Technology Acceptance Model (TAM), gathering feedback from 27 first-year students through a structured survey. In addition to these qualitative insights, quantitative measures such as memory (RAM) usage were evaluated to determine the efficiency and scalability of the tool under varying usage conditions.}

\onecolumn \maketitle \normalsize \setcounter{footnote}{0} \vfill

\section{\uppercase{Introduction}}
\label{sec:introduction}

Students' rapidly growing interest in learning programming has caused an increased adoption of APASs to manage and evaluate student exercises efficiently~\cite{tarek2022review}. 
An important aspect concerning the effectiveness of these systems is the ability to write code without downloading and installing an IDE on the students' computer \cite{horvath2018web} \cite{espana2017analyzing}. With the absence of an online IDE, APASs, in such cases, often require students to use a version control system (VCS), like \textit{GIT}, to submit their code. This is because APASs commonly use update events triggered by the VCS to execute a continuous integration (CI) pipeline and then automatically run pre-defined test cases on the students' code. The results of these tests are then used for assessment and are provided to the students as feedback.

One example of an APAS is the interactive learning platform \textit{Artemis}~\cite{krusche2018artemis} from the Technical University of Munich, which evaluates and assesses students' programming exercises and provides feedback to the students. Artemis includes an online editor, but students can also submit their programming assignments via a GIT repository. Automated tests are then carried out on the students' code in the repository, and the results are returned.

To avoid students also having to learn a VCS while learning programming, some APASs have introduced built-in online editors, allowing students to complete exercises directly within the platform \cite{krusche2018artemis} \cite{mekterovic2020building}. However, these online editors typically lack the comprehensive set of features commonly found in a modern desktop IDE. For example, syntax highlighting, auto-completion, and an integrated terminal are often missing. This gap in functionality can slow down the learning process and increase the entry barriers for students new to programming~\cite{leisner2019good}. An online IDE embedded in an APAS can support students in independently identifying issues and correcting errors in beginner programming. Native IDEs also support students with those features, but often student access to these native IDEs cannot be guaranteed due to the students' different backgrounds, resources, and technical equipment. Therefore, a feature-rich online IDE that can be accessed within a browser is a good way to ensure that every student has the same conditions and equal opportunities. The need for SE tools integrated into learning platforms, such as an online IDE for APAS, has also been identified in previous studies and underlines the need for development~\cite{meissner2024seet}.

To address these challenges, we found that an online IDE should at least incorporate several key features like syntax highlighting, auto-completion, and syntax error highlighting. Additionally, students should be able to compile and run code directly in the APAS without submitting their code, therefore triggering the execution of the test cases and leading to a significant overhead. 

While one might argue that this excessive tooling could inhibit the memorization of basic syntax, we believe this concern does not apply to programming education. Programming syntax is highly language-specific and does not necessarily translate to an understanding of the broader concepts essential for programming. For instance, focusing on learning the syntax of a single language by rote could lead to difficulties when transitioning to other languages. Instead, our position is that tools like syntax highlighting and code completion enable students to focus on solving problems and developing algorithms, which are fundamental skills in programming. These tools help students identify errors and improve their code quality, ultimately strengthening their conceptual understanding rather than detracting from it.

The primary goal of this work is to further investigate and address the mentioned challenges by exploring the following research questions:

\begin{itemize}[leftmargin=1em, itemsep=-1mm]
\item \textbf{RQ1:} What are the primary challenges and necessary architectural considerations for developing a feature-rich online IDE for APASs? 
\item \textbf{RQ2:} What are students' perceptions of the usability and usefulness of the new online IDE?
\end{itemize}

We conducted a comprehensive study involving several key steps to address these research questions. First, we gathered and defined key requirements for an online IDE in an educational context. Based on these requirements, we designed and implemented a prototypical open-source solution to be integrated with an APAS. Finally, we evaluated the implementation by conducting user studies to assess its effectiveness and gather student feedback.

Our evaluation shows that the implemented online IDE effectively addresses the identified challenges and is positively received by students, enhancing their learning experience.
Through this study, we aim to contribute to the field of programming education by providing insights into developing and integrating an effective online IDE within an APAS and presenting how the students perceive and utilize the tool.

The remainder of this paper is structured as follows: Section~\ref{sec:relwork} provides an overview of related work. Section~\ref{sec:requirements} elaborates on the requirements an online IDE should fulfill. Section~\ref{sec:implementation} then presents details about the prototypical implementation of the online IDE, which we evaluate in Section~\ref{sec:eval}. In Section~\ref{sec:disc}, we present the evaluation results and a respective discussion. Section~\ref{sec:threats} outlines potential threats to validity, and we conclude in Section~\ref{sec:conc}, reflecting on the broader implications of this work.

\section{\uppercase{Related Work}}
\label{sec:relwork}

Web-based IDEs are gaining popularity with both open-source and commercially hosted options becoming more widely available. For instance, \cite{Wu_2011} introduced \textit{CEclipse}, an online IDE designed for cloud-based programming. With CEclipse, the authors have focused on implementing an online IDE with functions similar to Eclipse's. However, since 2011, Eclipse has been developed further with new helpful functions, rendering CEclipse obsolete. In addition, the authors have mainly focused on program analysis issues to ensure security and analyze program behavior to improve programmer efficiency. In contrast, the approach presented in this paper focuses on a high-performance online IDE designed explicitly for APASs.

More recently, \cite{trachsler2018webtigerjython} unveiled \textit{WebTigerJython}, a web-based programming IDE for Python, featuring a clean interface and a textual code editor. However, the author focused on computer science courses at Swiss schools in his master's thesis. The requirements for this online IDE differed from ours, as the author had to ensure programming on a tablet and, therefore, only implemented basic functionalities for developing and executing Python programs in the browser. Compared to our work, the thesis did not focus on integrating APASs.

\cite{pawelczak2014virtual} implemented \textit{Virtual-C} to teach C programming with visualization tools to help students understand programming concepts. Their work integrated live coding functions in the lectures for programming assignments and self-learning functions. However, the authors of Virtual-C only focused on a programming environment for a C-programming course. Therefore, the general applicability in other programming languages is questionable. Also, since the paper was published in 2014, the enhancements to now common IDE features are not integrated into Virtual-C, making it outdated.

\cite{bigman2021pearprogram} presented \textit{PearProgram}, a learning tool for introductory computer science students. Their work focuses on pair programming activities in a remote learning setting. While PearProgram provides a code editor, its features are limited and focus on teaching the principles of pair programming rather than on programming itself, as we focus on in this paper.

Similarly to PearProgram, \textit{CodeHelper} by \cite{liu2021codehelper} is a lightweight IDE facilitating online pair programming in C++ courses. While CodeHelper also targets pair programming in education, the authors focused more on implementing and integrating CodeHelper but did not evaluate their approach. Furthermore, since the tool was created focusing on C++ courses, the general applicability in other programming languages remains uncertain.

\cite{tran2013interactive} implemented an online IDE called \textit{IDEOL} that leverages cloud computing to facilitate real-time collaborative coding and code execution. By shifting the heavy IDE cores to the cloud, IDEOL offers benefits such as work mobility, device independence, and lower local resource consumption. However, its features only focus on creating an interactive and responsive environment where real-time guidance, communication, and collaboration can be delivered.

Another notable web-based solution is \textit{MOCSIDE}, which offers a coding environment with auto-grading capabilities \cite{barlow2021mocside}. However, there is no documentation of the supported IDE features in its editor, and the open-source GitHub project seems not to exist anymore. 

The inclusion of commercial options like \textit{Repl.it}\footnote{\url{https://replit.com}}, \textit{GitHub Codespaces}\footnote{\url{https://github.com/features/codespaces}}, and \textit{Gitpod}\footnote{\url{https://www.gitpod.io}}, reveals an even broader spectrum of web-based IDEs with extensive feature support.
In a more detailed analysis of commercial online IDEs, we found that most of them use the open source editor \textit{Eclipse Theia}\footnote{\url{https://github.com/eclipse-theia/theia}}. However, there are few studies on the integration of Theia in APASs. An analysis of its performance in Section \ref{sec:implementation} also revealed significant concerns about using Theia in APASs.

Several APASs also introduced basic online editors to assist students with their submissions and general coding experiences. Notable platforms include \textit{Edgar} \cite{mekterovic2020building}, \textit{Fitchfork} \cite{pieterse2013automated}, \textit{CodeRunner} \cite{lobb2016coderunner}, \textit{CodeBoard} \cite{truniger2023codeboard}, \textit{CodeOcean} \cite{staubitz2016codeocean}, \textit{PythonTutor} \cite{guo2013online} and \textit{Artemis} \cite{krusche2018artemis}. However, these platforms typically only offer basic IDE features, which come out of the box by including editors like: \textit{ACE}\footnote{\url{https://ace.c9.io}}, \textit{Monaco Editor}\footnote{\url{https://github.com/microsoft/monaco-editor}} and \textit{CodeMirror}\footnote{\url{https://codemirror.net}}, therefore, lacking many of the more advanced IDE features, like advanced code-completion, syntax error highlighting or an integrated terminal or debugger, which is normally available in other native IDEs or commercial online IDEs.

Through the analysis of related work, we found that, up until now, there has still been a noticeable gap in the current capabilities provided by online IDEs used in APASs. While existing commercial online IDEs generally perform well in providing platforms for coding, they largely do not focus on teaching programming skills or evaluating coding exercises in an educational context. On the other hand, APASs do provide support for teaching coding and the assessment of programming exercises. Still, they often lack comprehensive IDE features such as advanced code completion, syntax error highlighting, and execution capabilities commonly found in commercial online IDEs. Thus, this study aims to bridge this gap by explicitly designing an online IDE for APASs and its prototypical implementation to evaluate the approach. 

\section{\uppercase{Requirements for Online IDEs in Programming Education}}
\label{sec:requirements}
To create an online IDE tailored to APASs, we must first define the relevant requirements that must be fulfilled. 
Thus, we used an anonymous online questionnaire to conduct a quantitative and qualitative analysis of students' requirements.

\subsection{Requirements Engineering Process}

To identify relevant requirements for online IDEs in APASs, we conducted an in-depth analysis of state-of-the-art desktop IDEs such as \textit{Visual Studio}\footnote{\url{https://code.visualstudio.com/}}, \textit{IntelliJ IDEA}\footnote{\url{https://www.jetbrains.com/idea/}}, and \textit{Eclipse}\footnote{\url{https://www.eclipse.org/downloads/}}. This analysis aimed to identify key features that contribute to their effectiveness and popularity among developers.

In a second step, to ensure that all requirements for an IDE in the educational context are covered, we asked 48 APAS users from three different Austrian universities: (1) Paris Lodron University Salzburg, (2) Johannes Kepler University Linz, and (3) University of Innsbruck to evaluate and rate the importance of the collected IDE features and, if needed, to provide additional features they perceive as important for an online IDE.

The survey respondents were predominantly first-semester students (n=35), with three participants each in their second and third semesters. One participant was in their fifth semester, one in the sixth, and one in the seventh semester. Four respondents did not specify their semester. 

The link to the survey containing the following questions was displayed on the login page of the Artemis platform: 

\begin{itemize}
    \setlength\itemsep{0.1em}
    \item \textbf{P1}: In which semester are you currently enrolled?
    \item \textbf{P2}: In how many university courses (including current ones) have you used Artemis?
    \item \textbf{P3}: How well do you know GIT commands and functionality? (Single choice regarding five options ranging from "not at all familiar" to "very familiar")
    \item \textbf{P4}: What tool did you primarily use to solve programming exercises in Artemis? (Single choice regarding five options ranging from "online editor only" to "local IDE only")
    \item Rate the importance of the following IDE features (Single choice on a five-point scale ranging from "Very Important" to "Very Unimportant")
    \begin{itemize}
        \item \textbf{F0}: Syntax Highlighting - uses colors to differentiate various components of the source code.
        \item \textbf{F1}: Debugger / Breakpoints - enables pausing code execution at specific lines to inspect variables and program state.
        \item \textbf{F2}: Auto-completion / Code Suggestions - offers possible code completions or suggestions based on the context of the typed code.
        \item \textbf{F3}: Syntax Error Highlighting - marks parts of the code that may lead to syntax errors, aiding in debugging.
        \item \textbf{F4}: Brace Matching Highlighting - highlights corresponding braces in the code, helping to visualize scope boundaries.
        \item \textbf{F5}: Key Shortcuts - includes shortcuts for common actions like copying, pasting, cutting, and searching.
        \item \textbf{F6}: Compiler / Interpreter (Run Button) - allows you to compile and run code without executing tests.
        \item \textbf{F7}: Dedicated Shell Console - provides access to a console where you can execute commands and run code manually with custom options.
    \end{itemize}
    \item \textbf{O1}: Do you have any additional suggestions for improving the code editor?
    \item \textbf{O2}: Are there any other critical features a code editor should have to better support your coding needs?
\end{itemize}

Additionally, for each feature question, we displayed an image of the feature when used in a traditional IDE to make sure that all students knew what the features were about. 

\subsection{Gathered Requirements}

The preliminary questions P1-P3 revealed that the majority is not familiar with \textit{GIT} commands (29\% - "Not at all familiar", 27\% - "Slightly familiar", 17\% "Neutral", 23\% "Moderately familiar" and only 4\% "Very familiar"). However, more than half of the students (58\%) use mostly local IDEs and GIT to solve the exercises, indicating that the available online editor is not feature-rich enough to complete an exercise there adequately.

Based on the feature questions (F1-F7) and mapping the answer possibility "Very Important" to $1$ and "Very Unimportant" to $5$, we were able to order the requirements by importance. The results are shown in Table~\ref{tab:mean_importance_ide_features}.

\begin{table}[h!]
    \centering
    \caption{Mean Importance of IDE Features (Ordered)}
    \label{tab:mean_importance_ide_features}
    \begin{tabular}{| m{4cm} | R{2cm} |}
        \hline
        \textbf{Feature} & \textbf{Mean Importance} \\ \hline
        Syntax Highlighting & 1.56 \\ \hline
        Syntax Error Highlighting & 1.69 \\ \hline
        Compiler / Interpreter (Run Button) & 2.03 \\ \hline
        Auto-completion / Code Suggestions & 2.06 \\ \hline
        Debugger / Breakpoints & 2.08 \\ \hline
        Brace Matching Highlighting & 2.19 \\ \hline
        Dedicated Shell Console & 2.44 \\ \hline
        Key Shortcuts & 2.47 \\ \hline
    \end{tabular}
\end{table}

Additionally, the open questions (O1 and O2) revealed that students value the following features: 

\begin{itemize}[leftmargin=1em, itemsep=-1mm]
    \item \textbf{Presentation Mode:} The IDE should include a presentation mode that rearranges the UI for better display of solutions in a classroom setting.
    \item \textbf{Integrated Console Output Pane:} To efficiently monitor the code output, the IDE should incorporate a console output pane within its interface.
    \item \textbf{Resizable Editor's Pane:} The IDE should allow users to resize the editor's pane, offering flexibility in managing the workspace according to their preferences.
    \item \textbf{Mobile Device Compatibility:} The IDE should be responsive for mobile devices, ensuring accessibility and usability on various devices.
    \item \textbf{Multiple File Handling:} It should support opening and working with multiple files simultaneously, allowing for more complex project development.
\end{itemize}

Additionally, two requirements, based on the setup at the University of Innsbruck, include:

\begin{itemize}[leftmargin=1em, itemsep=0em]
    \item \textbf{Compatibility with the APAS Artemis:} The IDE must be fully compatible with the Artemis learning platform, ensuring seamless integration and interaction to provide a cohesive user experience.
    \item \textbf{Low Resource Usage:} Given observed peaks of more than 500 weekly active users in the past three years on the Artemis platform, it is essential that the new IDE solution uses a limited amount of memory and is capable of scaling resources on-demand to handle exam situations. 
\end{itemize}

In conclusion, the requirements for the online IDE encompass both features to improve usability, along with scalability considerations, reflecting the needs of an educational environment with varying levels of user engagement. 

\section{\uppercase{Reference Implementation}}
\label{sec:implementation}

After gathering key requirements, the focus shifted towards evaluating existing solutions for the potential integration into Artemis. For this phase, we leveraged results from a previous study, where we conducted a systematic tool review of online IDEs \cite{frankfordrequirements}. As part of this, we assessed and extracted the set of supported features and programming languages and evaluated the availability of these tools as open-source solutions to ensure the source code is accessible and enables easy integration into APASs. The main findings were that existing online editors in APASs only support partial feature sets. The Artemis code editor, for example, already fulfills the following requirements: (1)~Syntax Highlighting, (2)~Syntax Error Highlighting, (3)~Multiple File Handling, and (4)~Low Resource Usage. However, commercial online IDEs, like Gitpod and Codespaces, support all IDE features as defined in the requirements specification. Still, we cannot access the commercial options' code bases, making integration impossible. 

As discussed in Section~\ref{sec:relwork}, we found that most of the online IDEs use the \textit{Eclipse Theia} editor, which is publicly available and open-source. Consequently, integrating Theia into the Artemis platform was considered a viable option. However, a closer examination of its memory usage revealed that Theia consumes too many resources, given that Artemis should be able to serve hundreds of users simultaneously. This increase in memory consumption, particularly during exam periods, where peaks are expected, raised concerns about Theia's feasibility. Even without Java plugins, the containerized Theia IDE exhibited a significantly larger memory footprint (approximately 135 MB) than Artemis' base Java Docker container. Adding Java plugins, essential for Java IntelliSense features, almost doubled the memory requirement to about 265 MB.

This observation led to the conclusion that Theia, in its entirety, is not a feasible solution for online IDEs in the educational sector where resources are limited. To mitigate this issue, the focus shifted to enhancing Artemis' native code editor.

One reason Theia has high memory requirements is that it integrates a language server handling the language server protocol (LSP), which is responsible for the IDE features directly within the IDE. Because of this, we decided to extract this component and use one language server for multiple user sessions, resulting in a significant decrease in memory usage. We were able to additionally implement a load-balancing mechanism, which always selects the language server with the least workload because the language server offers a set of key performance metrics like (1)~Average Load, (2)~CPU Usage, (3)~Total Memory, (4)~Free Memory and (5)~Number of Active Sessions.
As a result, we were able to address the resource concerns effectively. In the following Figure \ref{fig:ide_comparison}, you can see a comparison of the Artemis online editor's memory usage without IDE features, with IDE features (named online IDE), and then the reference, which is the Theia online IDE.

\begin{figure}[htbp]
    \centering
    \includegraphics[width=1\columnwidth]{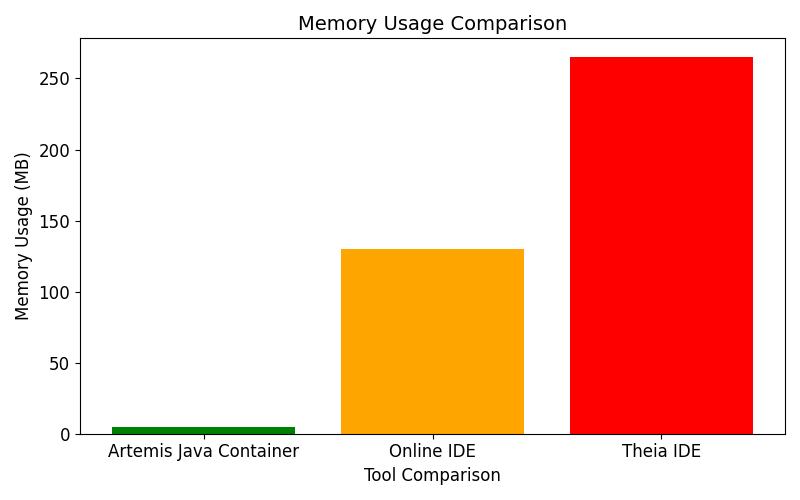}
    \caption{RAM comparison of the different IDEs.}
    \label{fig:ide_comparison}
\end{figure}

The VCS service in Artemis is also integral to our solution, serving as the central element for storing and distributing exercise data between Artemis and the LSP servers. This centralized VCS ensures consistent synchronization across instances. If a user session is closed, it can be reopened on a different LSP server, with the central VCS maintaining data coherence.

Incorporating data cloning for each user session into the LSP server's workflow inevitably leads to an escalation in disk space usage. This necessitates regular cleanups to eliminate data that is only temporarily utilized. We have integrated an automated cleanup system within each LSP server to streamline this process. This system checks every five minutes to identify inactive sessions -- defined as those with no LSP messages or file-modifying activity in the past 60 minutes. Subsequently, it removes them along with their corresponding files.

Another key requirement, the integrated terminal, introduces various security challenges due to the potential execution of arbitrary code and the risk of users trying to access other files by traversing the directories. This threat has been mitigated by ensuring that a new Docker container is started for each integrated terminal, and therefore, the code runs in a more or less controlled environment. 

Terminal sessions are initiated only upon user request to conserve resources and prevent unnecessary server load rather than starting automatically.

In addition to cleaning exercise data, we have also implemented a process for managing Docker containers spawned during sessions. While idle containers generally have minimal impact on resource usage and memory footprint, accumulating such containers could lead to inefficient resource utilization. This cleanup process is synchronized with the termination of the related web socket connection to improve resource management efficiency. Consequently, when a connection is closed or disrupted, and a container is no longer in use, its resources are released.

Analyzing the interaction between the new online IDE and the LSP server, we observed a substantial exchange of WebSocket messages during user sessions. This high communication volume is inherent to the LSP and its language servers, which need real-time updates with every user interaction, such as typing or hovering over code elements, to provide enhanced IDE features.
Despite the relatively small size of JSON-RPC messages, the sheer frequency and volume can impose a substantial load on the server when multiplied by the concurrent user count. 

To address this challenge, we have employed custom debouncing techniques within Artemis. Debouncing is a method that throttles the rate of operations, in this case, message transmissions. It works by postponing message dispatch until a certain period of inactivity has passed. For operations such as code lenses, documentation on hover, or code completion, we introduced a delay of one second, while file change actions were assigned a two-second delay.

\begin{figure}[htbp]
    \centering
    \includegraphics[width=1\columnwidth]{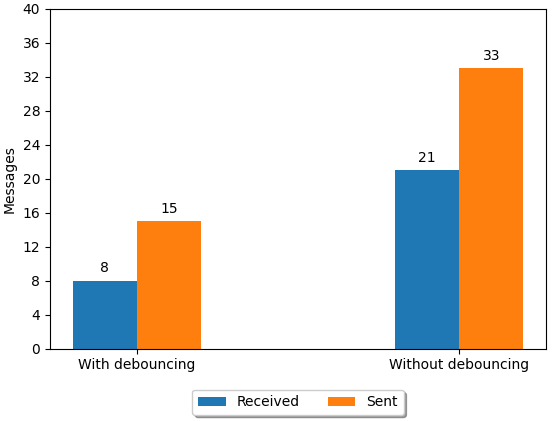}
    \caption{Impact of debouncing on the number of received and sent messages.}
    \label{fig:debouncing}
\end{figure}

The impact of this debouncing approach was evaluated through an experimental comparison. The experiment involved a user typing "Hello world!" in a Java exercise file on Artemis and measuring the message traffic with and without debouncing. 
The results, illustrated in Figure \ref{fig:debouncing}, indicate a significant reduction in messages sent. An approximate 43\% decrease in messages sent when debouncing was applied was achieved without negatively impacting user experience. This demonstrates the effectiveness of this technique in mitigating performance issues in an online IDE environment.

An additional 'run' button enhances the user interface of the web IDE, offering an automated execution of predefined command sequences with a single click. This sequence is designed to terminate any ongoing processes, clear the terminal, navigate to the relevant directory, and execute language-specific commands, such as \texttt{\footnotesize mvn clean test} for Java (Maven), or \texttt{\footnotesize pytest} for Python. This functionality replaces the need for a 'stop' button by allowing users to terminate and restart terminal sessions at will, thus addressing potential session freezes or faults.

\begin{figure*}[htbp]
    \centering
    \includegraphics[width=.7\textwidth]{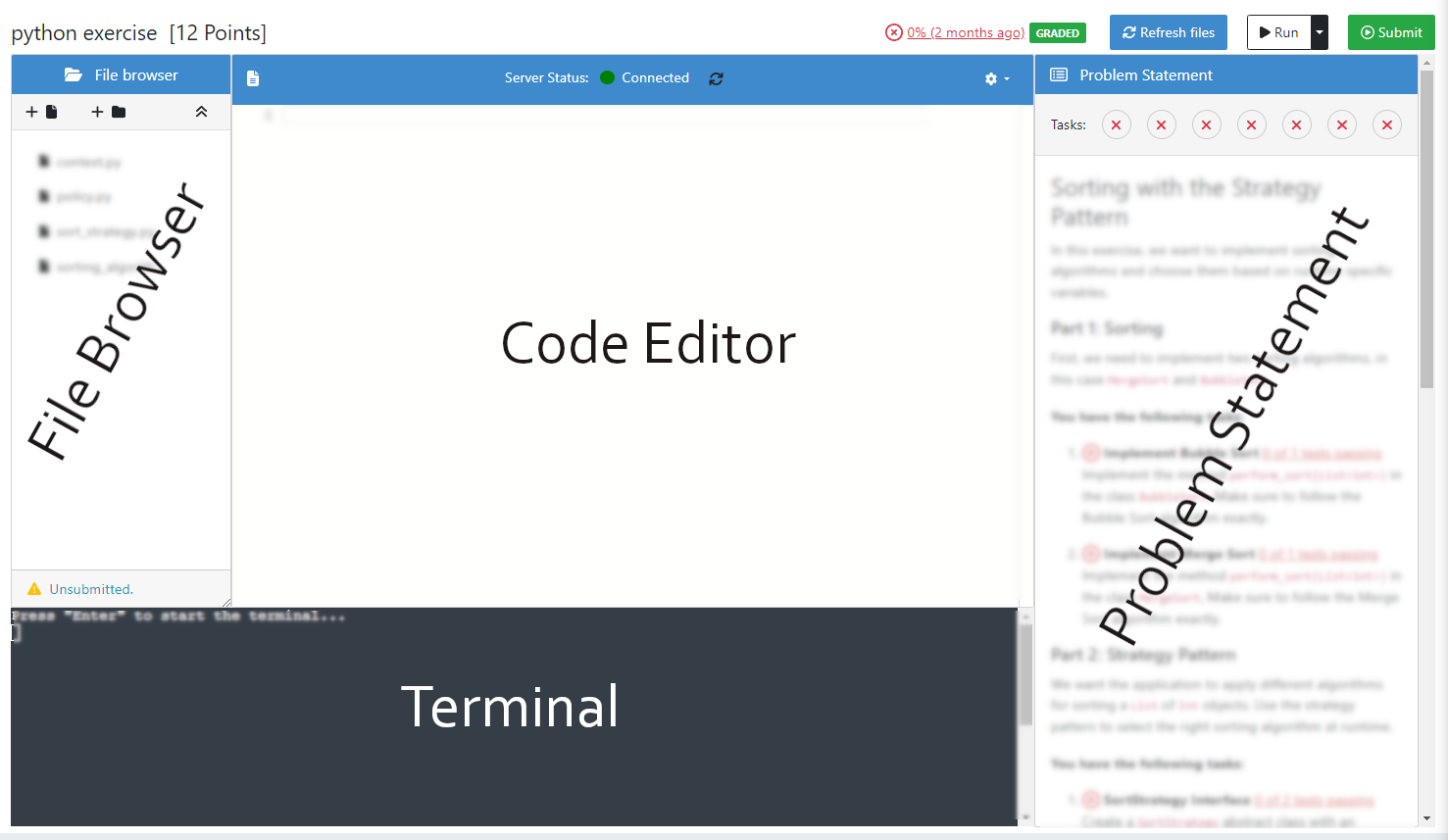}
    \caption{Overview of the new online IDE in Artemis.}
    \label{fig:ide}
\end{figure*}

The online IDE now comprises four main components: a (1)~file browser, a (2)~code editor, a (3)~problem statement viewer, and a (4)~terminal interface. Each component is designed to offer specific functionalities, as shown in the IDE overview in Figure~\ref{fig:ide}.
Currently, the LSP server solution supports Java, Python, and C exercises, representing the three main programming languages taught in courses at the University of Innsbruck.

\begin{findingsBox}
\textbf{Main Findings for RQ1}:
\begin{itemize}[leftmargin=1em, itemsep=-1mm]
    \item \textbf{Memory Efficiency:} Connecting multiple user sessions to a single LSP server significantly reduces memory usage, making the online IDE scalable for many concurrent users.
    \item \textbf{Load Balancing:} Implementing load-balancing mechanisms distributes workload evenly across servers, ensuring system stability during peak usage times like exams.
    \item \textbf{Security Measures:} Using restricted Docker containers for the integrated terminal addresses security concerns by isolating user code execution from the host system.
    \item \textbf{Optimized Communication:} Custom debouncing techniques reduce the communication load between clients and servers, enhancing performance without degrading the user experience.
\end{itemize}
\end{findingsBox}

\vspace{-3mm}

\section{\uppercase{Evaluation}}
\label{sec:eval}

In the following, we present the evaluation results of the new online IDE. We selected students from the "Introduction to Programming" tutorial, a compulsory course of the Bachelor's degree in Computer Science curriculum at the University of Innsbruck. This tutorial teaches first-year students the fundamentals of the C programming language. We received a total of n=27 valid responses. 

Prior to the survey, we introduced the students to the prototype of the online IDE that had been implemented. Afterward, the students were tasked with solving the Pascal's Triangle exercise \cite{hinz1992pascal} using only the online IDE. They had one week to complete the task.
In the next course, we allotted approximately 15 minutes for students to complete the online questionnaire.

\subsection{Study Design}
\label{studydesign}
We used the Technology Acceptance Model (TAM) to evaluate the prototype's usability.
TAM is a widely used research model in empirical user studies to determine the actual use of a technology based on the perceived usefulness (PU) and perceived ease of use (PEOU) \cite{chau1996empirical}. 
Additionally, for PU and PEOU we also evaluated four editor properties (CEE questions). This approach allowed us to systematically explore the users' perceptions concerning the perceived performance of code submissions, the quality of compilation feedback, the intuitiveness of the user interface, and their overall impression of the features provided \cite{davis1989perceived}. 

All questions used a 7-point Likert Scale from \emph{`Strongly Disagree'} (-3) to \emph{`Strongly Agree'} (3).
Regarding the PEOU, the following questions were asked:

\begin{itemize}[leftmargin=1em, itemsep=-1mm]
    \item \textbf{PEOU0:} Learning to operate the code editor is easy for me.
    \item \textbf{PEOU1:} I find it easy to get the code editor to do what I want it to do.
    \item \textbf{PEOU2:} My interaction with the code editor is clear and understandable.
    \item \textbf{PEOU3:} I find the code editor flexible to interact with.
    \item \textbf{PEOU4:} It is easy for me to become skillful at using the code editor.
    \item \textbf{PEOU5:} I find the code editor easy to use.
\end{itemize}

The following PU questions were asked:

\begin{itemize}[leftmargin=1em, itemsep=-1mm]
    \item \textbf{PU0:} Using the code editor in my tasks enables me to accomplish tasks more quickly.
    \item \textbf{PU1:} Using the code editor improves my task performance.
    \item \textbf{PU2:} Using the code editor in my tasks increases my productivity.
    \item \textbf{PU3:} Using the code editor enhances my effectiveness in the tasks.
    \item \textbf{PU4:} Using the code editor makes it easier to do my tasks.
    \item \textbf{PU5:} I find the code editor useful in my tasks.
\end{itemize}

The survey also included an explanation and screenshots for each of the properties that were additionally questioned. For these questions, a 5-point scale ranging from `Excellent' (1) to `Very poor' (5) has been used:

\begin{itemize}[leftmargin=1em, itemsep=-1mm]
    \item \textbf{CEE0: Submission / Testing performance:} The time it takes to receive feedback after submitting the exercise.
    \item \textbf{CEE1: Compilation / Testing feedback:} The feedback provided from the submission of an exercise.
    \item \textbf{CEE2: Editor’s user interface (UI):} The way the editor is structured and displayed.
    \item \textbf{CEE3: Editor’s features/functionalities:} The set of functionalities and features provided by the editor.
\end{itemize}

In addition to the TAM-based evaluation, we asked open questions again and measured the memory usage of the implemented online IDE. These measurements were critical to assess the IDE's performance efficiency, especially regarding resource consumption. This provided valuable insights into the IDE's scalability, an important requirement in the educational sector as computing and financial resources are limited.

\subsection{Results}

\begin{figure*}[htbp]
    \centering
    \includegraphics[width=1\textwidth]{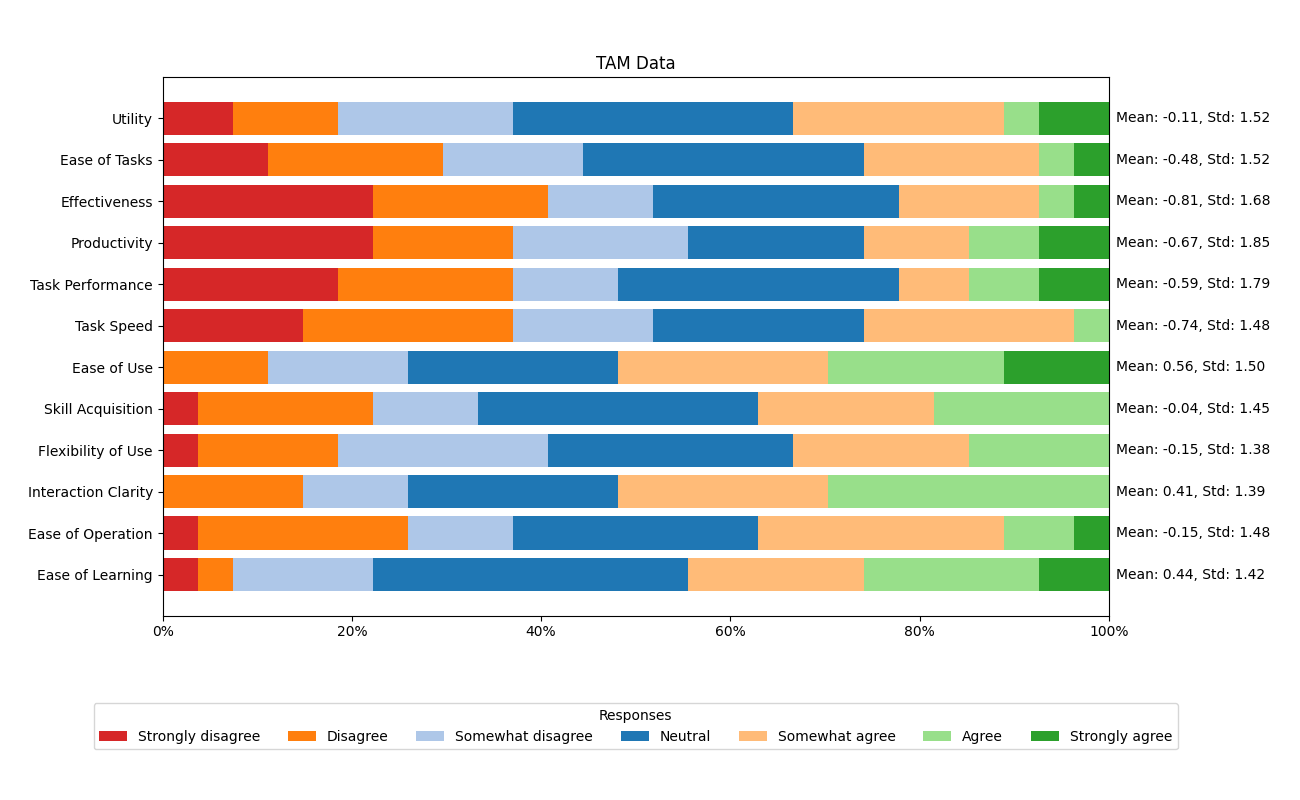} 
    \caption{Distribution of user responses to the TAM questionnaire.}
    \label{fig:TAM}
\end{figure*}

Figure~\ref{fig:TAM} presents findings from the TAM questionnaire to measure user attitudes towards the prototype IDE. Each bar represents the percentage of responses across a 7-point Likert scale, ranging from "Strongly disagree" to "Strongly agree" for various attributes.

Study participants generally agreed that the technology was easy to use and clear in interaction, and they found it reasonably easy to learn. \\ However, we observed a general disagreement regarding the utility, effectiveness, productivity, task performance, and task speed. High standard deviations suggest that some students appreciate the online IDE, but some seem to think the opposite.

As mentioned in Section \ref{studydesign}, we evaluated student perceptions of the online IDE based on four key aspects: (1)~Submission/Testing performance, (2)~Compilation/Testing feedback, (3)~Editor's user interface (UI), and (4)~Editor's features/functionalities.
'Compilation/Testing feedback' received the most favorable evaluation, with the highest mean score of 0.78 and a standard deviation of 0.83, indicating a generally positive user experience. The 'Editor user interface (UI)' scored well, with a mean of 0.67 and a standard deviation of 0.86, suggesting that users found the UI to be above average in terms of usability.
Conversely, 'Editor features/functionalities' and 'Submission/Testing performance' both recorded a mean score of 0.30 but with different levels of variance (standard deviations of 0.94 and 0.71, respectively). 

The open-ended feedback provided valuable insights into the students' perceptions and experiences with the online IDE, revealing both positive aspects and areas for improvement. Many participants expressed that the editor's utility is limited, with several preferring to use established IDEs like VSCode or CLion\footnote{\url{https://www.jetbrains.com/clion}} instead of the web-based environment. They highlighted that while the editor may be useful for introductory tasks, it is less suitable for advanced exercises, as they preferred external tools with more robust features.

Several usability concerns were also identified. Participants frequently mentioned the inability to resize or collapse UI panels, which disrupted their workflow, and requested more customization options, such as a dark mode and additional color schemes. Another common request was enhanced code assistance, including a more 'intelligent' auto-completion and better visible error highlighting. Some noted that the prototype's current implementation of these features was either too slow or inconsistent.

Despite these criticisms, some students appreciated the web IDE, citing its simplicity as a useful tool for beginners. Positive remarks included its utility for first-year students who may not yet be familiar with other development environments. Furthermore, feedback on compilation and testing functionality was generally positive, though there were isolated cases where users experienced discrepancies between local and web-based compilation results.

Users made specific feature requests, such as adding a 'vim mode,' implementing automated bracket closure, and improving the reporting of test results, particularly for failed tests. These suggestions indicate a desire for more advanced, developer-friendly features that align with those available in traditional IDEs.

While the open feedback highlights the current limitations of the web IDE, it also emphasizes the potential for enhancing its functionalities and user interface to better support novice and experienced users.

\begin{figure}[htbp]
\centering
\includegraphics[width=0.91\columnwidth]{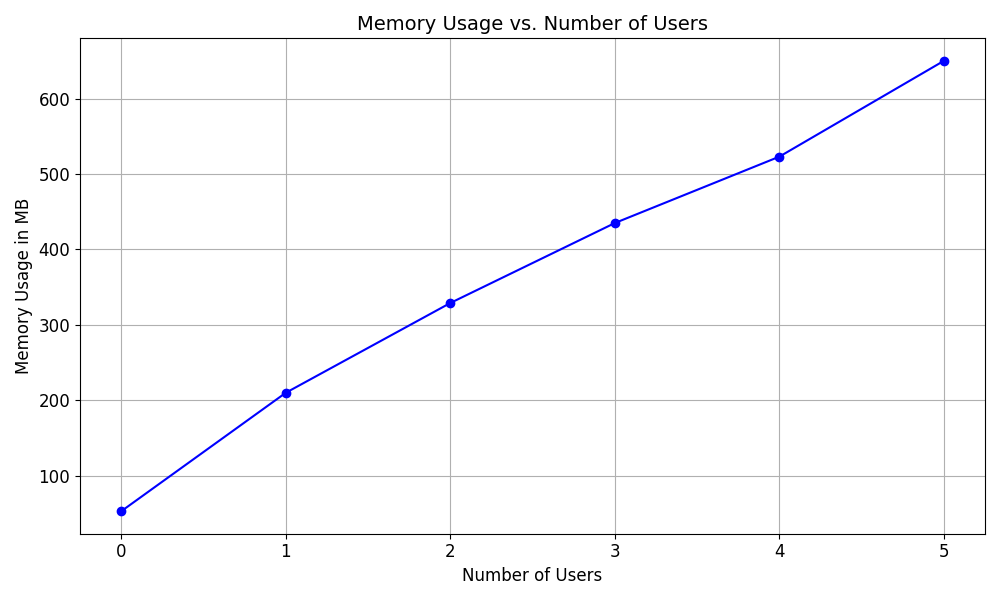}
\caption{Function of the online IDE's memory usage by the number of users.}
\label{fig:plotMemory}
\end{figure}

Figure \ref{fig:plotMemory} shows the memory usage of the prototype. This plot reveals a linear relationship between the number of users and the memory usage of the IDE. As the user count increased from 0 to 5, the memory consumption rose steadily from 53~MB to 650 MB. This pattern underscores the IDE's predictable resource consumption and suggests that the IDE demonstrates a capacity for scalability, with memory usage increasing by approximately 130~MB for each additional user. 

\begin{findingsBox}
\textbf{Main Findings for RQ2}:
\begin{itemize}[leftmargin=1em, itemsep=-1mm]
    \item \textbf{Perceived Ease of Use (PEOU):} Students generally found the online IDE easy to learn and use, with clear and understandable interactions.
    \item \textbf{Mixed Perceived Usefulness (PU):} While the IDE was appreciated for introductory tasks, many students felt it lacked utility for advanced exercises, leading to mixed opinions on its effectiveness and productivity enhancement.
    \item \textbf{Suitability for Beginners:} The simplicity of the web IDE was appreciated by first-year students who are not yet familiar with other development environments.
    \item \textbf{Areas for Improvement:} Students requested enhanced code assistance features and advanced functionalities, especially debugging tools or a 'vim mode'.
\end{itemize}
\end{findingsBox}

\section{\uppercase{Discussion}}
\label{sec:disc}

Developing and integrating an online IDE into an APAS such as Artemis presents both challenges and opportunities. This research aimed to address two primary research questions, leading to a series of significant findings that contribute to the field of programming education.

With the first research question, we focused on identifying the key challenges in implementing an online IDE for APASs. One main challenge we uncovered was the high memory consumption associated with comprehensive online IDE platforms like Theia\footnote{\url{https://github.com/eclipse-theia/theia}}. One Theia container needs approximately 260 MB of RAM, which would be prohibitively expensive in the educational domain, with peak usage of more than 500 parallel users. However, we were able to effectively mitigate this issue by connecting multiple editors to a single LSP server and by introducing load-balancing, resulting in a significant memory reduction by around 50\% improving both economic feasibility while still offering a good user experience during peak usage times such as exams. Moreover, we effectively addressed security concerns by implementing a Docker environment for the integrated terminal.

Another key challenge was the high communication load between the online IDE and LSP servers, which we addressed by implementing custom debouncing techniques in our prototypical online IDE. This optimization reduced communication intensity, further enhancing the system's efficiency. Additionally, regular session cleanup emerged as crucial for maintaining system performance, emphasizing the need for efficient resource management strategies to develop online IDEs for educational purposes.

The investigation into user perceptions (RQ2) revealed mixed responses regarding the new online IDE's utility, effectiveness, and productivity.  While most users found the technology generally easy to use and learn, some disagreed on its impact on task performance and speed. These findings suggest that while the new IDE has improved certain aspects of the user experience, there are still areas that require further improvement.

The open feedback offers valuable guidance for further enhancing the IDE. Several participants suggested that while they appreciated the simplicity of the web IDE for learning basic concepts, they would like to see features that support more advanced coding tasks, similar to those offered by established tools like VSCode or CLion. This indicates an opportunity for the IDE to evolve beyond its current form by incorporating advanced features such as a fully integrated debugger, better code completion, and detailed error highlighting. By expanding the feature set, the online IDE could become a more attractive option for experienced users who prefer working within a single, consistent environment.

The feedback also included innovative suggestions, such as adding a 'vim mode' and a quick-save feature (e.g., mapped to Ctrl-S), illustrating how users are eager to see features that enhance efficiency and familiarity. These ideas present opportunities for future development. By responding to these user-driven insights, the IDE has the potential to bridge the gap between a beginner-friendly platform and a powerful development tool.

The positive reception of the compilation and testing functionalities demonstrates that the core features of the online IDE are functioning well, providing users with reliable tools for checking their code. 

In summary, the open feedback highlights the potential of the online IDE as a flexible and scalable solution that, with further development, can accommodate a wide range of user needs. The suggestions from users provide a clear roadmap for future improvements, emphasizing the importance of striking a balance between simplicity for beginners and advanced functionality for more experienced users. By incorporating these insights, the IDE could become an even more effective tool for programming education.

\section{\uppercase{Limitations}}
\label{sec:threats}

In this paper, we mainly considered the following four categories of validity, also used by \cite{wohlin2012experimentation}: (1) Construct validity, (2) Reliability, (3) Internal validity, and (4) External validity.

\subsection{Construct Validity}

Construct validity in this study refers to how  accurately the research design and methods capture the key challenges of implementing an online IDE and the subsequent user perceptions. The challenges identified, such as memory consumption, load balancing, and security issues, were derived from the technical implementation process and user feedback. While these challenges represent common issues faced in software development, especially in educational settings, they may not encompass all possible hurdles that could emerge in different contexts or with other technologies.

The study's construct validity may also be influenced by the specific features and functionalities implemented in the prototype. The chosen set of features aimed to address the most critical needs of programming students. However, this selection process might not fully capture the entire spectrum of useful IDE features, especially those that could be more relevant in different programming languages, paradigms, or more advanced levels of study.

\subsection{Reliability}

The reliability of this study is significantly improved by the open-source nature of the prototypical implementation of the online IDE within the Artemis platform. By making the source code and the questionnaire results publicly available on Zenodo\footnote{\url{https://doi.org/10.5281/zenodo.13959741}}, the study allows verification and replication by the broader community, thereby improving the transparency and trustworthiness of the implementation process. This open-source approach ensures that every aspect of the IDE development, from the initial design decisions to the specific coding practices, is accessible for review and reuse. 

\subsection{External Validity}

A potential limitation of this study comes from integrating the online IDE exclusively within the Artemis platform. While Artemis is an APAS with functionalities common to many APASs, this exclusive integration may affect the generalizability of the findings. However, the architectural and functional principles applied in developing the online IDE, such as using LSP servers for multiple user sessions and Docker containers for sandboxing, are platform-agnostic. These principles could be adapted for integration into other educational settings, suggesting that our findings have relevance beyond the Artemis environment.

The participant pool of students from a single "Introduction to Programming" course for the second survey also limits external validity. With 27 respondents, the sample size, while sufficient for a preliminary investigation, limits the ability to conduct comprehensive quantitative analyses and make broad generalizations. 

\subsection{Internal Validity}

The internal validity of this study is based on the careful design and execution aimed at minimizing external influences on the variables investigated. Key strategies to improve internal validity included a uniform and well-tested questionnaire (TAM) and leveraging objective metrics for system evaluation like memory usage. 

\section{\uppercase{Conclusion}}
\label{sec:conc}

This research presents an innovative concept for an online IDE designed for programming education. We developed and implemented a prototype to evaluate the concept.
By providing a feature-rich online IDE directly within an educational platform, we can offer a more engaging and effective learning experience for students beginning their programming journey.

The prototypical implementation addresses critical technical challenges (RQ1), such as high memory consumption and security issues, by integrating multiple editors with a single LSP server and leveraging Docker containers for sandboxed execution environments, and the open-source nature of the presented solution ensures transparency and encourages community involvement for further refinement. The evaluation (RQ2) suggests that while the new online IDE is user-friendly and easy to learn, there is room for improvement in demonstrating its utility and impact on task performance. The mixed opinions on utility and productivity highlight the need for further enhancements to increase the practical benefits of the IDE in real-world tasks. 

We plan to refine the IDE for future work by incorporating user feedback to enhance features and functionalities. This includes implementing advanced tools such as an integrated debugger, improving code completion, and optimizing performance to reduce load times and improve reliability. We also aim to conduct more extensive user studies across different educational institutions and platforms to validate and generalize our findings.

\balance
\bibliographystyle{apalike}
{\small
\bibliography{example}}

\end{document}